\documentclass[11pt,twoside]{article}
\usepackage{index}
\usepackage{graphicx}

\begin{document}

\begin{center}

\begin{LARGE}
Formation and Self-Assembly of Coherent Quantum Dots: Some
Thermodynamic Aspects
\end{LARGE}

\vspace{1cm}

{\bf Jos\'e Emilio Prieto}

Centro de Microan\'alisis de Materiales and \\
Instituto Universitario de
Ciencia de Materiales ``Nicol\'as Cabrera", \\
Universidad Aut\'onoma de Madrid, 28049 Madrid, Spain

{\bf Elka Korutcheva}

Dpto. de F\'{\i}sica Fundamental, UNED, Senda del Rey 9, 28080
Madrid, Spain

{\bf Ivan Markov}

Institute of Physical Chemistry, Bulgarian Academy of Sciences, 1113
Sofia, Bulgaria

\end{center}

\begin{abstract}
Quantum dots have promising properties for optoelectronic
applications. They can be grown free of dislocations in highly
mismatched epitaxy in the coherent Stranski-Krastanov mode. In this
chapter, some thermodynamic aspects related to the wetting in the
growth and self-assembly of three-dimensional (3D) coherent islands
are studied using an energy minimization scheme in a
1+1-dimensional atomic model with anharmonic interactions. The
conditions for equilibrium between the different phases are
discussed. It is found that the thermodynamic driving force for
3D-cluster formation is the reduced adhesion of the islands to the
wetting layer at their edges. In agreement with experimental
observations, for values of the lattice mismatch larger than a
critical misfit, a critical island size for the 2D-3D transition is
found. Beyond it, monolayer islands become unstable against bilayer
ones. Compressed coherent overlayers show a greater tendency to
clustering than expanded ones. The transition to 3D islands takes
place through a series of intermediate stable states with
thicknesses discretely increasing in monolayer steps. Special
emphasis is made on the analysis of the critical misfit.
Additionally, the effect of neighbouring islands mediated through a
deformable wetting layer is considered. The degree of wetting of the
substrate by a given island depends on the size and shape
distributions of the neighbouring islands. Implications for the
self-assembled growth of quantum dots are discussed.
\end{abstract}

\section{Introduction}

The fabrication of ordered arrays of self-assembled,
three-dimensional (3D) nanosized islands is the subject of intense
research in recent times due to their possible applications in the
fields of nanomagnetism and optoelectronics.\cite{Politi} To obtain
semiconducting ``quantum dots'' that can be efficiently used as
lasers and light-emitting diodes, it is convenient to grow islands
which are {\em coherent} with the atomic layers underneath, i.e.,
that do not contain misfit dislocations (MDs). One promising
approach to the fabrication of such structures is to exploit the
``coherent Stranski-Krastanov (SK) growth'' in the epitaxy of highly
mismatched materials. This case is different from the ``classical''
SK growth in which the lattice mismatch is accommodated by MDs at
the interface with the wetting layer.\cite{Matt} The term
``coherent'' SK growth has been coined for this case of formation of
3D islands that are strained to fit the underlying strained wetting
layer of the same material at the interface but are largely
strain-free near their top and side walls.\cite{Eagle,Vit} It has
been recently reported in a series of systems, e.g.  Ge on
Si,\cite{Eagle,Letanh,Mo,Aumann,Voigt1,Legoues,Ross,Kaestner} InAs
on GaAs,\cite{Moison,Heitz,Kobayashi,Yamaguchi,Joyce,Polimeni}
InGaAs on GaAs,\cite{Leo,Guha,Sneider,Temmyo} InP on
In$_{0.5}$Ga$_{0.5}$P,\cite{Carlsson} that coherently strained
(dislocation-free) 3D islands grow on the wetting layer. In all
these cases, the lattice misfit is positive and very large (4.2, 7.2
and about 3.8~\% for Ge/Si, InAs/GaAs and InP/In$_{0.5}$Ga$_{0.5}$P,
respectively) for semicoductor materials, which are characterized by
directional and brittle chemical bonds. A possible (partial)
exception might be the system PbSe/PbTe(111), where the misfit is
negative (-5.5~\%) and the overlayer is expanded.\cite{Pinc}

The growth of thin epitaxial films takes place usually in conditions
far from equilibrium, particularly when using deposition methods in
vacuum like molecular beam epitaxy. Nevertheless, thermodynamic
considerations are a necessary step to understand many aspects
of the growth process. Of particular interest for the case of the
coherent SK growth is the thermodynamic driving force (TDF)
responsible for the 3D islanding. The reason is that we have in fact
the growth of $A$ on strained $A$, the coherent 3D islands being
strained to the same degree as the wetting layer. A naive conclusion
from this would be that the energy of the interfacial boundary
between the 3D islands and the wetting layer is equal to zero.
Although this energy is certainly expected to be small compared to
the free energy of the crystal faces involved~\cite{Politi}, it
cannot be neglected since this would imply that the islands wet
completely the wetting layer, ruling out 3D islanding from a
thermodynamic point of view~\cite{Rudy}.

The need for a thermodynamic anal\-y\-sis arises also from the
experimental observation of a critical misfit for coherent 3D
islanding,\cite{Leo,Pinc,Xie,Walther} and the si\-mul\-ta\-neous
presence of islands of different thicknesses which vary by one
monolayer (ML).\cite{Rudra,Col} The existence of a critical misfit,
as well as of stable two-, three- or four-ML-thick islands, do not
follow from the tradeoff~\cite{Politi,Jerry}
\begin{eqnarray}\label{energy}
\Delta E \approx C^{\prime }\gamma V^{2/3} -
C^{\prime\prime}\varepsilon _{0}^{2}V
\end{eqnarray}
be\-tween the cost of the additional surface energy and the gain of
energy due to the elastic relaxation of the 3D islands relative to
the wetting layer ($V$, $\gamma $ and $\varepsilon _{0}$ are the
islands volume, the specific surface energy and the lattice misfit,
respectively, and $C^{\prime }$ and $C^{\prime\prime}$ are
constants).

For these reasons, in this chapter we first recall some simple
thermodynamic aspects of the epitaxial morphology based on the
traditional concept of wetting and consider the coherent SK growth
from this point of view. The basic ideas are then supported by
numerical calculations that make use of a simple model in $1+1$
dimensions with anharmonic interactions.

In order for the 3D islands to be useful for technical applications,
they have to be produced with high areal densities and with a high
degree of similarity and homogeneity. Experimental studies of arrays
of coherent 3D islands of highly mismatched se\-mi\-con\-duc\-tor
ma\-te\-ri\-als have shown surprisingly narrow size
distributions.\cite{Moison,Leo,Gru,Jia} This phenomenon, known as
self-assembly,\cite{Chris} is highly desirable as it guarantees a
spe\-ci\-fic optical wavelength of the ar\-ray of quantum dots.

In the second part of this chapter, we address the topic of
self-assembly. We consider no longer single isolated islands, but
groups of islands on top of a deformable wetting layer consisting of
several atomic planes that are allowed to relax in response to the
strain fields produced by the misfitting islands. In this way,
elastic interactions between the islands can be studied. We will
focus on the wetting of islands that are part of an array, i.e.,
that are surrounded by neighbouring islands. We study the effect of
the density of the array (the distance to the nearest neighbouring
islands), the size distribution (the difference in size of the
neighbouring islands) and the shape distribution (the slope of the
side walls of the neighbouring islands) on the wetting parameter
$\Phi $ of the considered island. The results are finally discussed
and compared to experiments.

\section{Wetting in the coherent SK growth}

We consider two different phases. The first is one which we can call
a mother or ambient phase. We can think of it as the vapour phase.
The second is a condensed phase and can consist either of a strained
planar film or of unstrained 3D crystals. A fundamental
thermodynamical result is that two phases are in equilibrium with
each other when their chemical potentials are equal. A transition
from one phase to a second one takes place when the chemical
potential of the second becomes smaller than that of the first. The
TDF for the transition is thus the difference of the chemical
potentials of both phases at the given pressure and temperature.
Therefore, the TDF which determines the occurrence of one or another
mechanism of epitaxial growth (i.e., the growth of a coherently
strained 2D layer or of 3D crystallites) is the difference

\begin{equation}
\Delta \mu =\mu (n)-\mu _{3D}^{0}
\end{equation}
of the chemical potential of the overlayer $\mu (n)$, which depends
on the average film thickness (measured in number $n$ of monolayers
counted from the interface), and the chemical potential, $\mu
_{3D}^{0}$, of the bulk 3D crystal of the same
material.\cite{Kern,Stoyan,Mark5}

The thickness dependence of the film chemical potential $\mu (n)$
originates from the thickness distribution of the misfit strain and
from the interaction between the deposit and the substrate, which
rapidly decreases with the distance from the interface and can
usually be neglected beyond several MLs.\cite{Kern,Stoyan,Mark5}

We can describe the essential phenomenology in terms of the wetting
considering a simple model of epitaxial growth of a crystal A on a
substrate B. At this stage, we do not include any lattice misfit
between the two materials but only a difference in the strength of
the chemical bonding.

As is well known, the wetting parameter which accounts for the
energetic influence of the crystal B on A is defined as (for a
review see Ref. \cite{Mar2})

\begin{equation}\label{phi}
\Phi  = 1 - \frac{E_{\rm AB}}{E_{\rm AA}},
\end{equation}
where $E_{\rm AA}$ and $E_{\rm AB}$ are the energies per atom
required to disjoin a half-crystal A from a like half-crystal A and
from an unlike half-crystal B, respectively.

The chemical potential of the bulk crystal A is given at zero
temperature by $ - \phi _{\rm AA}$, where $\phi _{\rm AA}$ is the
work required to detach an atom from the well known kink or
half-crystal position. This name is due to the fact that an atom at
this position is bound to a half-atomic row, a half-crystal plane
and a half-crystal block.\cite{Koss,Stran} In the case of a
monolayer-thick film of A on the surface of B, the chemical
potential of an atom A is given by the analogous term $ - \phi _{\rm
AB}$ whereby the underlying half-crystal block of A has been
replaced by a half-crystal block of B. Therefore we have

\begin{equation}\label{delta}
\Delta \mu  = \phi _{\rm AA} - \phi _{\rm AB}.
\end{equation}

In the simplest case of additivity of the energies of lateral and
vertical bonds, this difference reduces to $E_{\rm AA} - E_{\rm AB}$
as the lateral bondings cancel each other (it is easy to show that
Eq.~\ref{delta} is equivalent to the well known 3-$\sigma$ criterion
of Bauer~\cite{Bauer,Mar2}). Then $\Delta \mu $ is proportional to
$\Phi $:\cite{Mar2}

\begin{equation}
\label{dmueaaphi} \Delta \mu  = E_{\rm AA}\Phi.
\end{equation}

\begin{figure}[htb]
\center
\includegraphics*[width=8.0cm]{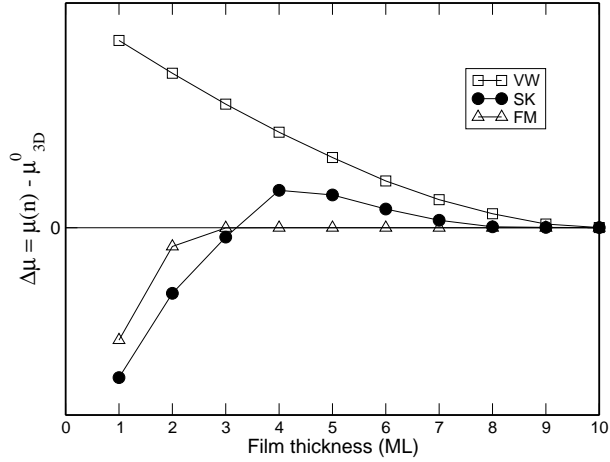}
\caption{\label{grmodes} \small Schematic dependence of the film
chemical pot\-ent\-ial on the film thickness in monolayers for the
three classical modes of growth: VW -- Volmer-Weber, SK --
Stranski-Krastanov, and FM -- Frank-van der Merwe.}
\end{figure}

It follows that it is in fact the wetting parameter $\Phi $ which
determines the mechanism of growth of A on B\cite{Rudy} as
illustrated by Figure~\ref{grmodes}. We first consider two limiting
cases. The Volmer-Weber (VW) growth of isolated 3D islands of A on
the surface of B takes place at any misfit $\varepsilon _0 $ and is
characterized by the incomplete wetting of the substrate ($0 < \Phi
< 1$). The value of $\Delta \mu$ is always positive and decreases
asymptotically to zero with increasing thickness. The opposite limit
is the consecutive formation of MLs of A on B in the Frank - van der
Merwe (FW) growth, with complete wetting ($\Phi \le 0$). For this
mode of growth is required that $\varepsilon _0 \approx 0$. In this
case, $\Delta \mu $ is always negative and tends asymptotically to
zero with increasing thickness.

Also shown in Figure~\ref{grmodes} is the intermediate and important
case of the SK mode of growth, which takes place when the atoms of
the deposit bind stronger to those of the substrate than to
themselves ($E_{\rm AB} > E_{\rm AA}$) but the lattice misfit
$\varepsilon_0 $ is substantial. The first condition causes a
complete wetting at the beginning of the growth ($\Phi < 0$) but the
second produces an accumulation of strain energy with increasing
film thickness which turns the balance towards $\Phi > 0$
(incomplete wetting) at some value of the thickness $n$. In the
classical SK growth, the introduction of MDs allows for a release of
strain energy through a structural relaxation of the film and to the
development of 3D islands.

It is instructive to consider also Fig.~\ref{grmodes} in terms of
equilibrium vapour pressures. Although the con\-nec\-tion with the
chemical potential is straightforward ($\mu  \propto \ln P$), this
gives a deeper insight into the problem. Thus, as long as $\mu (n) <
\mu _{\rm 3D}^{0}$ a thin planar film can be deposited at a vapour
pressure $P$ smaller than the equilibrium vapour pressure $P_{0}$ of
the bulk crystal, but larger than the equilibrium vapour pressure
$P_{1}$ of the first ML, i.e. $P_1 < P < P_0$. In other words, a
planar film can be deposited at {\it undersaturation} $\Delta \mu  =
kT\ln (P/P_0)$ with respect to the bulk crystal. The formation of 3D
islands [$\mu (n) > \mu _{\rm 3D}^{0}$] requires $P > P_0$, or a
{\it su\-per\-sa\-tu\-ra\-tion} with respect to the bulk crystal. We
thus conclude that the 3D islands and the wetting layer represent
nec\-es\-sar\-ily different phases and thus have different chemical
potentials. The reason is that the two phases are in
equ\-i\-lib\-ri\-um with the mother phase under different conditions
which never overlap.\cite{prtmrk1} The wetting layer can be in
equ\-i\-lib\-ri\-um only with an un\-der\-sa\-tu\-rated vapour phase
($P < P_0$), while small 3D islands can be in equilibrium only with
a su\-per\-sa\-tu\-ra\-ted vapour phase ($P > P_0$). The dividing
line is thus $\Delta \mu  = kT\ln (P/P_0) = 0$ at which the wetting
layer cannot grow thicker and the 3D islands cannot nucleate and
grow. Hence the wetting layer and the 3D islands can never be in
equilibrium with each other.

The derivative, $d\Delta E/dV$, of the energy of the 3D islands
relative to that of the wetting layer, gives the difference of the
chemical potentials of the wetting layer and the chemical potential
of the 3D islands. In other words, it represents the difference of
the supersaturations of the vapour phase with respect to the wetting
layer and the 3D islands. Transfer of material from the stable
wetting layer defined by $\mu _{\rm WL} < \mu _{\rm 3D}^{0}$ to the
3D islands is connected with an increase of the free energy of the
system and therefore is thermodynamically unfavoured. A planar film
thicker than the stable wetting layer is unstable and the excess of
the material must be transferred to the 3D islands if the necessary
thermal activation exists.

The Stranski-Krastanov morphology appears as a results of the
interplay of the film-substrate bonding, misfit strain and the
surface energies. A wetting layer with a thickness of the order of
the range of the interatomic forces is first formed (owing to the
interplay of the A-B interaction and the strain energy accumulation)
on top of which partially or completely relaxed 3D islands nucleate
and grow. The 3D islands and the thermodynamically stable wetting
layer represent necessarily different phases. If this were not the
case, the growth would continue by 2D layers. Thus we can consider
as a useful approximation to regard the 3D islanding on top of the
uniformly strained wetting layer as a Volmer-Weber growth. That
requires the mean adhesion of the atoms that belong to the base
plane of the 3D islands to the stable wetting layer to be smaller
than the cohesion between them. In other words, the wetting of the
underlying wetting layer by the 3D islands must be incomplete.
Otherwise, 3D islanding will not occur.\cite{Rudy} In the
Volmer-Weber growth the incomplete wetting is due mainly to the
difference in bonding ($E_{\rm AB} < E_{\rm AA}$), the supplementary
effect of the lattice misfit being usually smaller. In the classical
SK growth, the incomplete wetting is due to the energetic cost of
the array of MDs.

In view of this, the case of {\em coherent} SK growth seems
difficult to understand, since in the absence of MDs there is no
obvious difference between the atoms of the wetting layer and those
of the islands. If this is so, then there is no evident reason for the
consideration of two different phases, the wetting must be complete
and 3D islanding should not occur.\cite{Rudy}

\begin{figure}[htb]
\center
\includegraphics*[width=8.0cm]{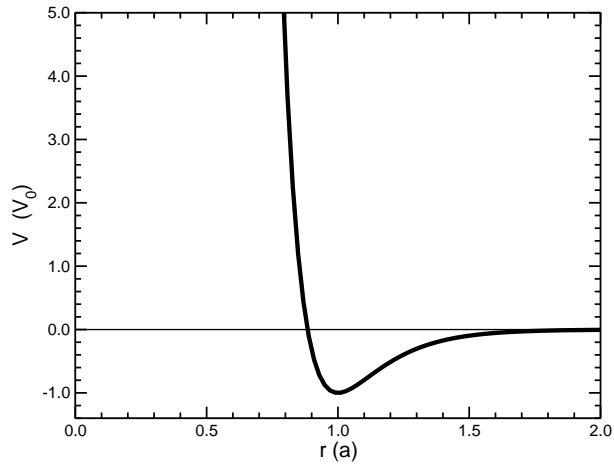}
\caption{\label{Vmorse} \small Graphical representation of the
atomic interaction potential, of the Morse type, used in the
calculations in this work.}
\end{figure}

It has been recently suggested that the TDF for the {\em coherent}
SK growth, where $E_{\rm AB} \approx E_{\rm AA}$, is the incomplete
wetting of the substrate by the islands.\cite{Kor} The origin of the
reduced adhesion are the displacements of the atoms near the island
edges from the bottoms of the corresponding potential troughs
provided by the wetting layer underneath. However, the
ap\-pro\-xi\-ma\-tion used by the authors, which is based on the 1D
model of Fren\-kel and Kontorova,\cite{Ratsch,Frenkel} is unable to
describe correctly the individual behaviour of atoms inside each
layer, since it assumes a potential with a period given by the
average of the separations of atoms (considered frozen) in the layer
underneath. Although this model gives qua\-li\-ta\-ti\-ve\-ly
reasonable results concerning the energy of the islands, it is
inadequate to calculate, in particular, the structure and energy of
the interfacial boundary between the wetting layer and the 3D
islands upon thickening of the latter. The model is further unable
to consider the effect of neighbouring islands on the wetting of a
given one.

\subsection{Model}

For the above-mentioned reasons, we have performed atomistic
calculations making use of a simple minimization procedure. The
atoms interact through a pair potential whose anharmonicity can be
varied by adjusting two constants $\mu $ and $\nu $ ($\mu  > \nu $)
that govern separately the repulsive and the attractive branches,
respectively,\cite{Mar}
\begin{eqnarray}\label{potent}
V(x) = V_{o}\Biggl[\frac{\nu }{\mu - \nu }e^{-\mu (x-a)} - \frac{\mu
}{\mu - \nu }e^{-\nu (x-a)}\Biggr],
\end{eqnarray}
where $a$ is the equilibrium atom separation. For $\mu  = 2\nu $ the
potential (\ref{potent}) turns into the familiar Morse form, which
has been used in the present work for the particular case $\nu = 6
$. This potential is shown in Figure~\ref{Vmorse}.

\begin{figure}[htb]
\center
\includegraphics*[width=8.0cm]{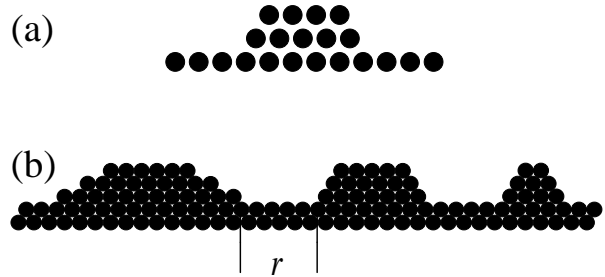}
\caption{\label{scheme} \small Schematic view of (a) an isolated
island on a rigid substrate with side angles of 60$^\circ$ and (b)
an array of islands on a wetting layer. The central island is
surrounded by two islands with different shapes and sizes. The
distance between neighbouring islands is given by~$r$.}
\end{figure}

Our programs calculate the interaction energy of all the atoms as
well as its gradient with respect to the atomic coordinates, i.e.
the forces. Relaxation of the system is performed by allowing the
atoms to displace in the direction of the gradient in an iterative
procedure until the forces fall below some negligible cutoff value.
The calculations for isolated islands were performed under the
assumption that the substrate (the wetting layer) is rigid. It was
carefully checked that removing this restriction, i.e., allowing
several layers below the islands to relax did not change
qualitatively the results.

We consider an atomistic model in $1+1$ dimensions which can be
regarded as a cross-section of the real $2+1$ case. An implicit
assumption is that the islands have a compact rather than a fractal
shape and that the lattice misfit is the same in both orthogonal
directions. The 3D islands are represented by linear chains of atoms
stacked one upon the other~\cite{Ratsch} as shown schematically in
Fig.~\ref{scheme}(a). The shape of the islands in our model is given
by the slope of the side walls.

In the case of arrays of islands, the restriction of a rigid
substrate must obviously be lifted in order for it to mediate
elastic interactions between the islands. The 1+1-dimensional array
is represented by a row of 3 (occasionally 5) islands on a wetting
layer consisting of several MLs which are allowed to relax. The
distance between two neighbouring islands is given by the number $r$
of vacant atomic positions between the ends of their base chains, as
illustrated by Fig.~\ref{scheme}(b). Periodic boundary conditions
are applied in the lateral direction.

For the sake of simplicity in the computational procedure, we
consider in our model a wetting layer that is in fact composed of
several MLs of the true wetting layer of the overlayer material A
plus several MLs of the unlike substrate material B. This
``composite wet\-ting layer" has the atom spacing of the substrate
material B as in the real case, but the atom bonding of the
overlayer material A. This underestimates somewhat the value of
$\Phi $, because the A-A bonding is necessarily weaker than the B-B
bonding in order for SK growth to be possible, but it does not
introduce a significant error since the energetic influence of the
substrate B is screened by the true wetting layer A.

In the present work, we will be primarily concerned with the
wetting. It is therefore important to define precisely the wetting
parameter~$\Phi$. In the coherent SK the relation between $\Delta
\mu $ and $\Phi $ is not as simple as that given by
Eq.~\ref{dmueaaphi}, because the bond energies are generally not
additive, the misfit strain is relaxed mostly near the side and top
walls and increasing the island thickness leads to larger
displacements of the edge atoms, as will be shown below. For these
reasons, in this work we define the wetting parameter $\Phi $ as the
difference of the interaction energies with the wetting layer of
misfitting and non-misfitting 3D islands.

\subsection{Atomistics at the interface}

\begin{figure}[htb]
\center
\includegraphics*[width=14.0cm]{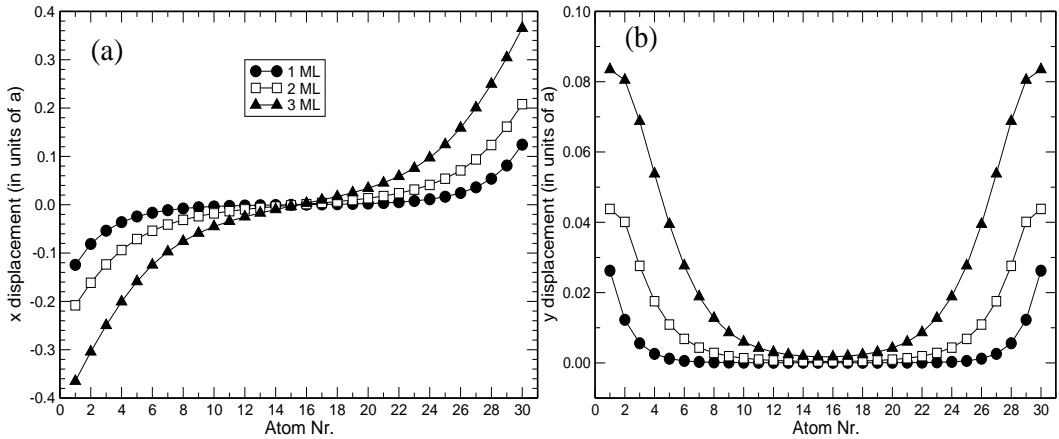}
\caption{\label{xycoh} \small Horizontal (a) and vertical (b)
displacements of the atoms of the base chain from the bottoms of the
potential troughs provided by the homogeneously strained wetting
layer for a misfit of 7.0~\%. The displacements are given in units
of $a$, the lattice parameter of the substrate and wetting layer,
for different island thicknesses. Islands of 30 atoms in the base
chain were considered.}
\end{figure}

We first consider the characteristics of the adhesion between the
{\em coherent} 3D islands and the wetting layer. Fig.~\ref{xycoh}(a)
shows the horizontal displacements of the atoms of the base chain.
They are measured from the bottoms of the corresponding potential
troughs of the wetting layer. It can be seen that the end atoms are
strongly displaced as in the model of Frenkel and
Kontorova,\cite{Frenkel} and of Frank and van der Merwe.\cite{FM}
Increasing the island height leads to larger displacements of the
end atoms. The reason is the effective increase of the strength of
the lateral interatomic bonding in the overlayer with greater
thickness as predicted by van der Merwe {\it et al.}\cite{Voltaire}
Fig.~\ref{xycoh}(b) shows the vertical displacements of the
base-chain atoms relative to the interplanar spacing between the MLs
belonging to the wetting layer. The vertical displacements are
obviously due to the climbing on top of the underlying atoms as a
result of the horizontal displacements. The thicker the island, the
larger the horizontal and, in turn, the vertical displacements. The
consequence is that the island loses contact with the underlying
wetting layer.

\begin{figure}[htb]
\center
\includegraphics*[width=14.0cm]{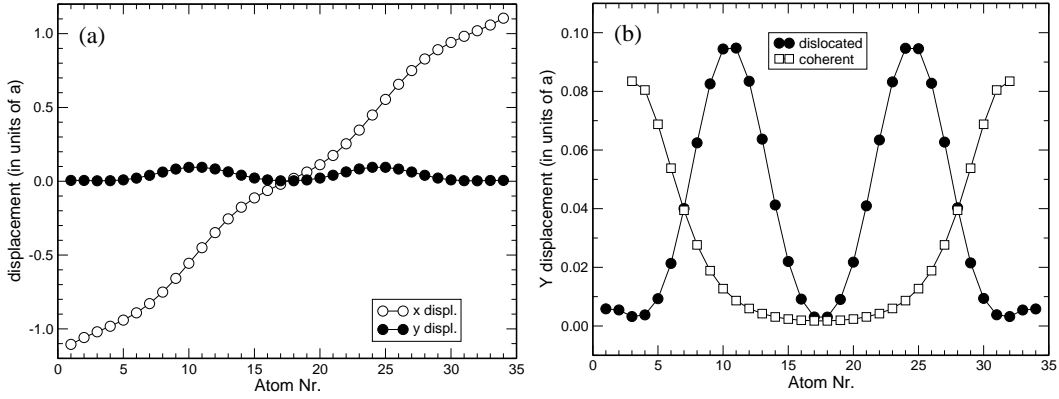}
\caption{\label{xycohdisl} \small (a): horizontal (x) and vertical
(y) displacements of the atoms of the base chain of a 3~ML-thick
island that contains two MDs. The island has a total amount of 99
atoms (34 in the base chain) and the lattice misfit is 7.0~\%. The
vertical displacements are shown in an enlarged scale in (b),
together with those of a coherent island of the same thickness and
30 atoms in the base chain.  Displacements are given in units of the
lattice parameter of the wetting layer and are measured from the
bottoms of the potential troughs provided by it.}
\end{figure}

It is instructive to compare these atomic displacements in
coherently strained islands with those found in the {\em classical}
SK growth. For this case, the interconnection between vertical and
horizontal displacements is shown in Fig.~\ref{xycohdisl}(a) for an
island containing two MDs. The horizontal displacements are now
larger in the cores of the MDs and so are also the vertical
displacements. Figure~\ref{xycohdisl}(b) is a further illustration
of the differences and resemblances between the classical and the
coherent SK modes. It shows the vertical displacements for both
types of 3D islands. The sizes of the base chain of the islands (30
and 34 atoms, respectively) are just below and above the critical
size for introduction of misfit dislocations at the given values of
the thickness (3~ML) and the lattice misfit (7.0~\%). As seen, in
both cases the 3D islands loose contact with the wetting layer. The
vertical displacements are largest at the chain's ends in the
coherent SK mode and around the dislocation cores in the classical
case, but the physics is essentially the same. In both cases we find
a reduction of the adhesion which leads to the 3D islanding.

\begin{figure}[htb]
\center
\includegraphics*[width=8.0cm]{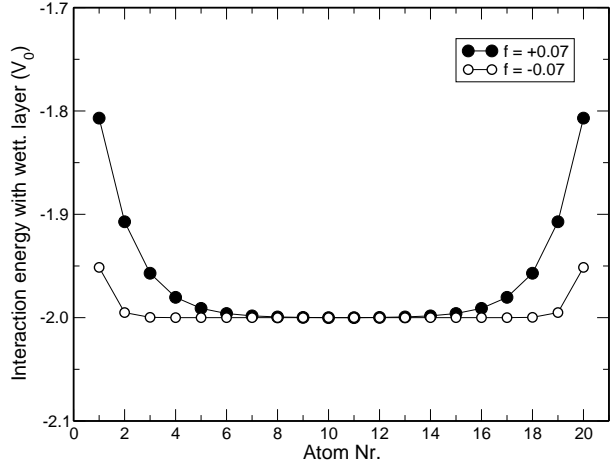}
\caption{\label{phivsn} \small Distribution of the energy (in units
of $V_{0}$) of first-neighbours interaction between the atoms of a
monolayer-high coherent island of 20 atoms and the underlying
wetting layer, for positive and negative misfits of absolute value
7.0~\%.}
\end{figure}

The power of our atomistic approach is that it allows us to study
the behaviour of the individual atoms that form an island. In order
to illustrate the effect of the atom displacements on the adhesion
of the separate atoms belonging to the island base chain, we have
plotted their energy of interaction with the underlying wetting
layer in Fig.~\ref{phivsn} for coherently strained islands with
values of the misfit of +7.0~\% and -7.0~\%. It can be seen that the
atoms that are closer to the island's edges adhere weaker to the
substrate. The influence of the potential anharmonicity is also
clearly demonstrated. Only one or two end atoms in the expanded
chain adhere weaker to the substrate whereas more than half of the
atoms at both ends do so in the compressed chain. The figure
demonstrates in fact the physical reason for the coherent SK mode
which is often overlooked in theoretical models. Moreover, it is a
clear evidence of why compressed rather than expanded overlayers
exhibit greater tendency to coherent SK growth.

\begin{figure}[htb]
\center
\includegraphics*[width=8.0cm]{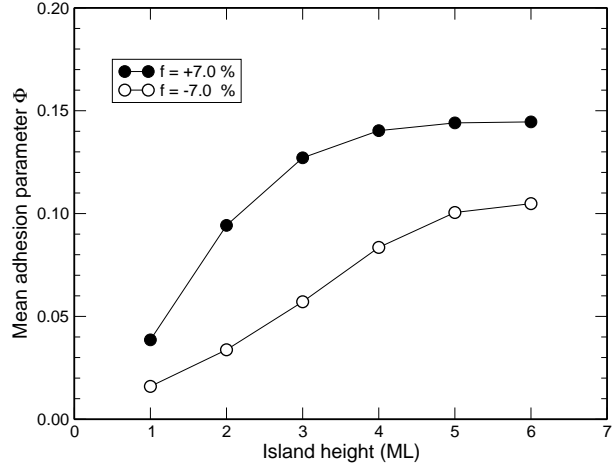}
\caption{\label{phivsh} \small Mean adhesion parameter $ \Phi $ as a
function of the islands' height in number of MLs for positive and
negative values of the misfit of absolute value of 7.0~\%. Coherent
islands of 14 atoms in the base chain were considered in the
calculations.}
\end{figure}

Another consequence of the atomic displacements shown in
Fig.~\ref{xycoh} is that increasing the islands thickness leads to a
weaker adhesion with the wetting layer and, in turn, to a
stabilization of the coherent 3D islands. This is demonstrated in
Fig.~\ref{phivsh}, which shows the dependence of the mean adhesion
parameter $ \Phi $ on the island height. As seen it saturates beyond
a thickness of about 5~ML. This effect has been predicted recently
by Korutcheva {\em et al.}\cite{Kor} by using the approximation
suggested by van der Merwe {\it et al}.\cite{Voltaire}

\begin{figure}[htb]
\center
\includegraphics*[width=8.0cm]{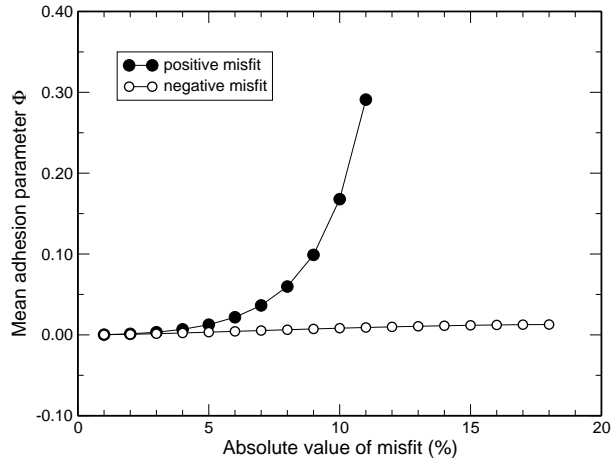}
\caption{\label{phivsf} \small Mean adhesion parameter as a function
of the lattice misfit for coherent, 1~ML-thick islands containing 20
atoms.  Data for both positive and negative misfits are shown in the
same quadrant for easier comparison.}
\end{figure}

The same tendency is demonstrated even more clearly in
Fig.~\ref{phivsf} where the mean adhesion parameter of coherent
monolayer-high islands is plotted as a function of the lattice
misfit in both cases of compressed and expanded overlayers. In fact
the figure shows the TDF $\Delta \mu $ for the transformation of
monolayer-high islands into 3D islands. The wetting parameter
increases very slowly and is nearly equal to zero up to about 12~\%
for expanded overlayers, whereas it increases steeply beyond a value
of the misfit of approximately 5~\% in compressed overlayers. As
will be shown below, this is in agreement with the misfit dependence
of the critical size (volume) for the mono-bilayer transformation to
occur.

An additional proof that the TDF for coherent 3D islanding is the
reduced adhesion is obtained by studying the energies of 2D and 3D
islands as a function of the total number of atoms or the volume. A
direct consequence of such a study is the existence of a critical
misfit beyond which coherent 3D islanding can take place. We compare
the energies per atom of mono- and multilayer high islands with
different thickness increasing discretely by one monolayer. One of
the reason for doing this is that islands with different thickness
have different adhesion parameters (see Fig.~\ref{phivsh}).
Moreover, as shown by Stoyanov and Markov, such an approach
originates from the classical concept of minimum of the surface
energy at a fixed volume.\cite{StMar} It is particularly applicable
to small crystallites formed on unlike surfaces where the crystal
height is a discrete rather than a continuous variable.

\begin{figure}[htb]
\center
\includegraphics*[width=14.0cm]{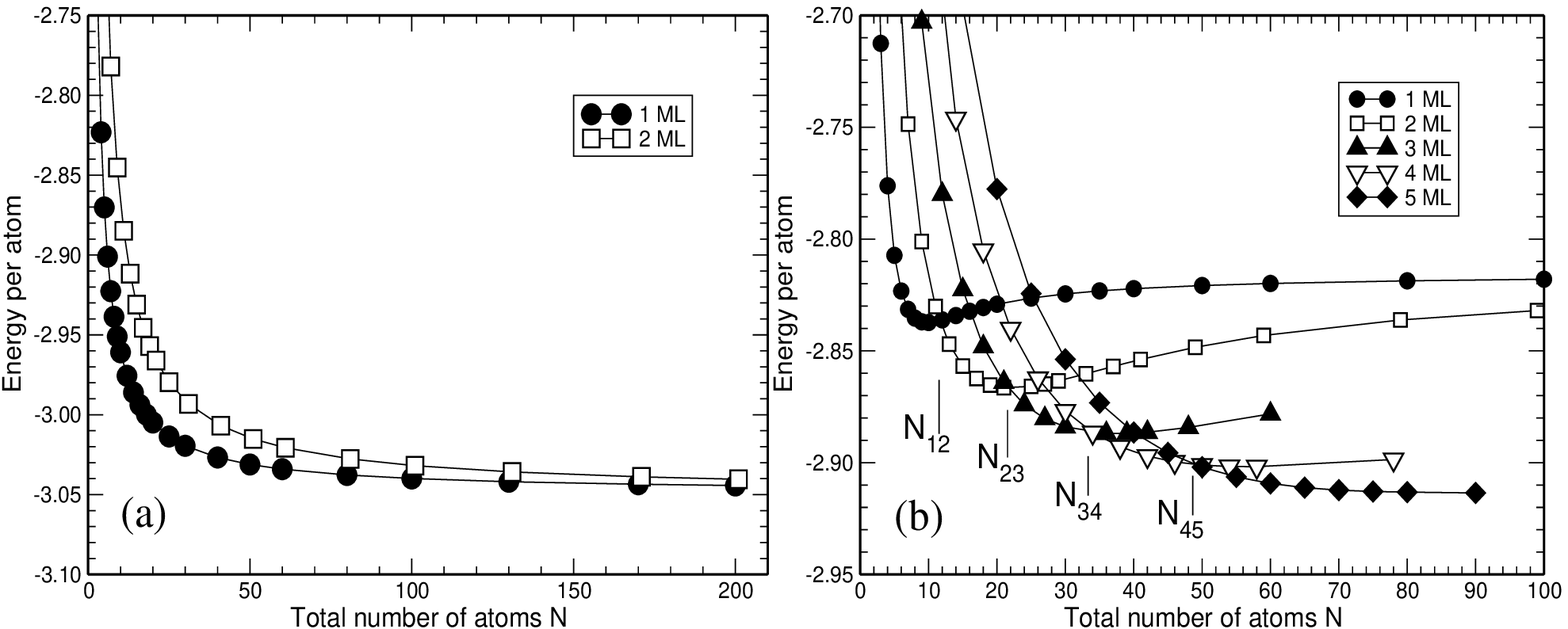}
\caption{\label{evsn} \small Dependence of the energy per atom, in
units of $V_0$, on the total number of atoms in compressed,
coherently strained islands of different thicknesses for two
different values of the misfit: (a) 3.0~\%, (b) 7.0~\%. The numbers
$N_{12}$, $N_{23}$, etc. give the limits of stability of monolayer,
bilayer... islands, respectively.}
\end{figure}

Fig.~\ref{evsn}(a) shows the energies per atom plotted as a function
of the total number of atoms in coherent monolayer- and bilayer-high
islands at $\varepsilon _{0} = 3.0~\%$. As seen, the monolayer
islands are always stable against the bilayer islands. This means
that thermodynamics does not favour coherent 3D islanding.
Monolayer-high islands will grow and coalesce until they cover the
whole surface. Misfit dislocations will be then introduced to
relieve the strain. The reason for this behaviour is clearly the
negligible value of the adhesion parameter as shown in
Fig.~\ref{phivsf}. Figure~\ref{evsn}(b) demonstrates the same
dependence (including also thicker islands) for a larger value of
the misfit $\varepsilon _{0} = 7.0~\%$, at which the adhesion
parameter has clearly a non-zero value. This time the behaviour is
completely different. Monolayer islands are stable against bilayer
islands only up to a critical volume $N_{12}$, the bilayer islands
are stable in turn against the trilayer islands up to a second
critical volume $N_{23}$, etc.\cite{corfu} This behaviour is
precisely the same as in the case of VW growth where the interatomic
forces $AA$ and $AB$ differ and the lattice misfit plays an
additional role.\cite{StMar} The same result (not shown) has been
obtained in the case of expanded overlayers ($\varepsilon _{0} < 0$)
with the only exception that monolayer-high islands are stable
against multilayer islands up to much larger absolute values of the
misfit.

\begin{figure}[htb]
\center
\includegraphics*[width=8.0cm]{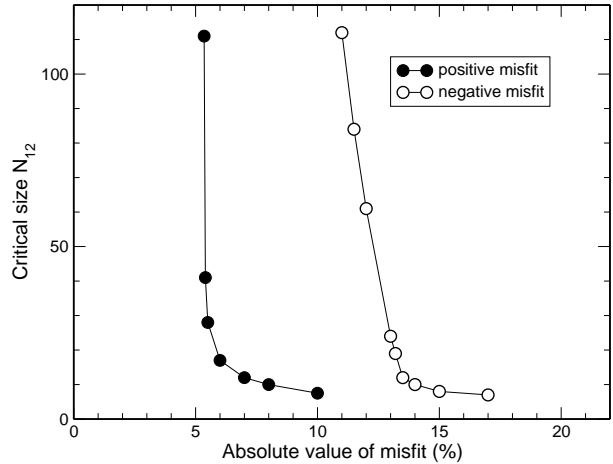}
\caption{\label{n12vsf} \small Misfit dependence of the critical
size $N_{12}$ (in number of atoms) for positive and negative values
of the lattice misfit. The curves are shown in one quadrant for
easier comparison.}
\end{figure}

Figure~\ref{evsn}(b) suggests that the mono-bilayer transformation
appears as the first step of the complete 2D-3D transformation. The
dependence of the critical size $N_{12}$ on the lattice misfit, as
shown in Fig.~\ref{n12vsf}, shows the existence of critical misfits
only beyond which the formation of multilayer islands can take
place. Below the critical misfit the monolayer-high islands are
stable irrespective of their size and the growth will continue in a
layer-by-layer mode until misfit dislocations are introduced at the
interface to relax the strain. The much larger absolute value of the
negative critical misfit is obviously due to the anharmonicity of
the atomic interactions. The weaker attractive interatomic forces
lead to smaller displacements, both lateral and vertical, of the end
atoms and in turn to stronger adhesion. Thus, a larger misfit is
required in order for 3D islanding to take place.

\section{Self-assembly}

\begin{figure}[htb]
\center
\includegraphics*[width=14.0cm]{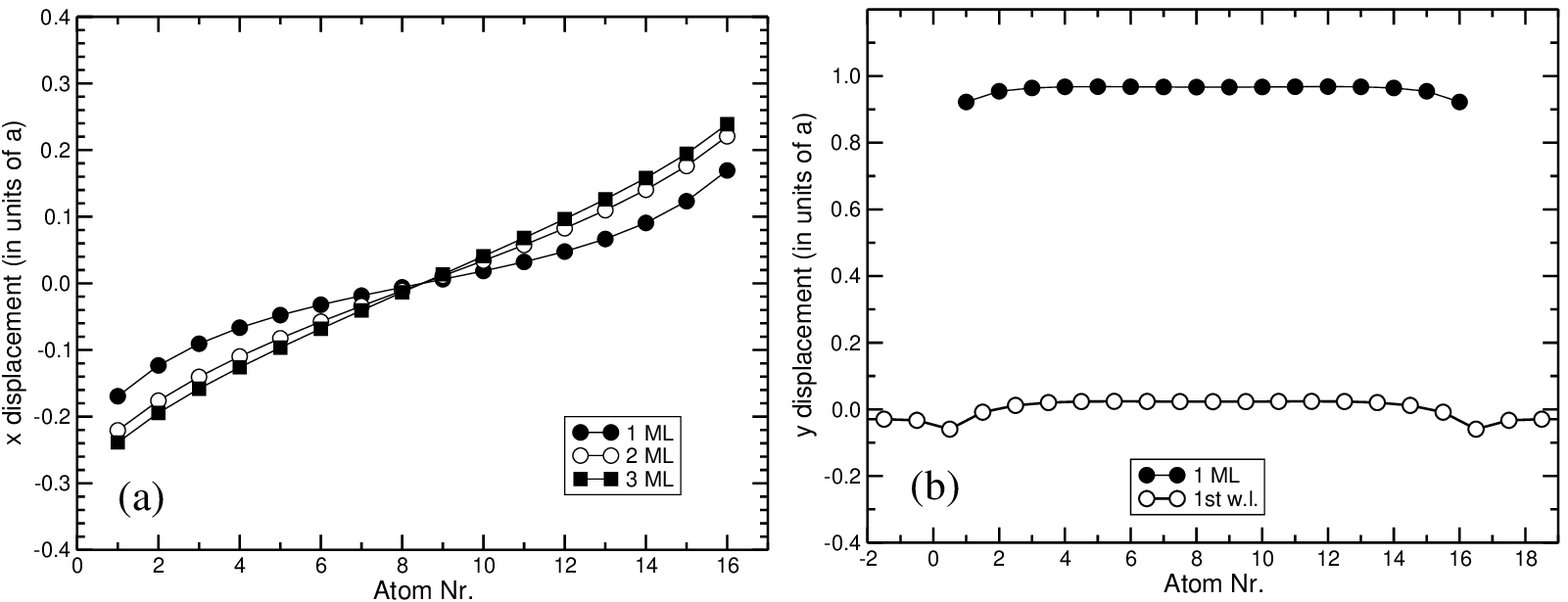}
\caption{\label{xydisplarr} \small (a): horizontal displacements of
the atoms of the base chain for a lattice misfit of 7.0~\% and the
given island thicknesses. The considered islands have 16 atoms in
their base chains and are located between two identical ones at a
distance of $r = 5$. The displacements are given in units of the
lattice parameter $a$ of the composite wetting layer, which consists
of 10 layers allowed to relax. (b): vertical displacements of the
atoms of the central island and of the first layer of the wetting
layer for the case of 1~ML-high islands.}
\end{figure}

The atomic displacements in an island that belongs to an array of
islands are shown in Fig.~\ref{xydisplarr}. Panel (a) shows the
horizontal displacements of the atoms of the base chain from the
bottoms of the potential troughs provided by the homogeneously
strained wetting layer for a misfit of 7.0~\%. The considered island
has two identical neighbours at both sides at a distance of $r = 5$.
This is the same behaviour as predicted by the one-dimensional model
of Frank and van der Merwe.\cite{FM} The horizontal displacements
increase with increasing island thickness (measured in number of
MLs) precisely as in the case of a rigid substrate and
non-interacting islands (see Fig.~\ref{xycoh}). But, in contrast to
this case, now the vertical displacements of the edge atoms of the
base chain of the islands and the underlying atoms of the uppermost
ML of the wetting layer are directed downwards, as shown in
Fig.~\ref{xydisplarr}(b). A similar result has been found by Lysenko
{\it et al.} in the case of homoepitaxial metal growth by means of a
computational method based on the tight-binding model.\cite{Lys}

\begin{figure}[htb]
\center
\includegraphics*[width=8.0cm]{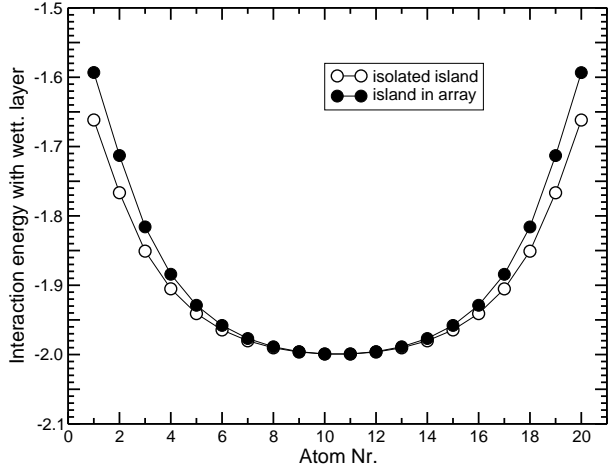}
\caption{\label{phivsiarr} \small Distribution of the interaction
energy (in units of $V_0$) between the atoms of the base chain of a
3~ML-high, coherent island of 20 atoms in the base chain, and the
underlying wetting layer, for a positive misfit of 8.0~\%. Full
circles correspond to an island separated by a distance $r = 5$ from
two identical neighbours, while the empty ones correspond to a
reference isolated island.}
\end{figure}

In spite of their downwards vertical displacements, the edge atoms
are again more weakly bound to the underlying wetting layer, as in
the rigid substrate case (Fig.~\ref{xycoh}). This is nicely
illustrated in Fig.~\ref{phivsiarr}. Shown in the same figure for
comparison is also an island without neighbours on the same relaxed
wetting layer. It can be seen in both cases that the edge atoms bind
weaker than the central atoms to the underlying wetting layer.
Compared to an isolated island, the edge atoms of an island in an
array adhere weaker to the substrate. Thus the essential physical
effect exerted by neighbouring islands on a give one is the
additional loss of contact of the latter with the substrate: the
wetting parameter must increase.

\begin{figure}[htb]
\center
\includegraphics*[width=8.0cm]{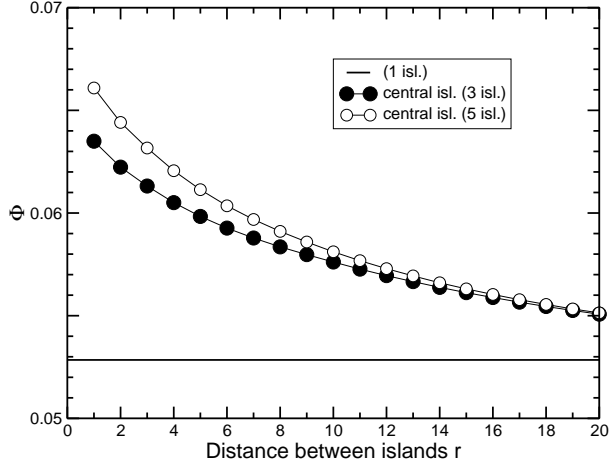}
\caption{\label{phivsrarr} \small Dependence of the wetting
parameter of the central island on the distance $r$ between the
islands. All islands are 3~ML high and have 20 atoms in the base
chain. The lattice misfit amounts to 7.0~\%. Results for arrays of 3
and 5 islands are given, as well as for a reference isolated
island.}
\end{figure}

The behaviour of the wetting parameter as a function of the density
of the array is illustrated in Fig.~\ref{phivsrarr}. As expected,
the wetting parameter increases with increasing array density. This
figure also allows us to estimate the size of the effect. At a
distance $r$=10 (about 3~nm), neighbours make $\Phi$ increase by
10~\%; this represents an effective decrease in adhesion $\Delta
E_{AB}$ of 0.10 $\Phi$ $E_{AA}$. For Ge/Si(100) (desorption energy
of Ge: 4~eV), this gives about 20~meV, a contribution of the same
order as the elastic energy per atom (40~meV in this
system~\cite{Mar2}) that can significantly affect the delicate
balance of the energies involved in the growth process: diffusion
barriers and surface/interface energies.

\begin{figure}[htb]
\center
\includegraphics*[width=14.0cm]{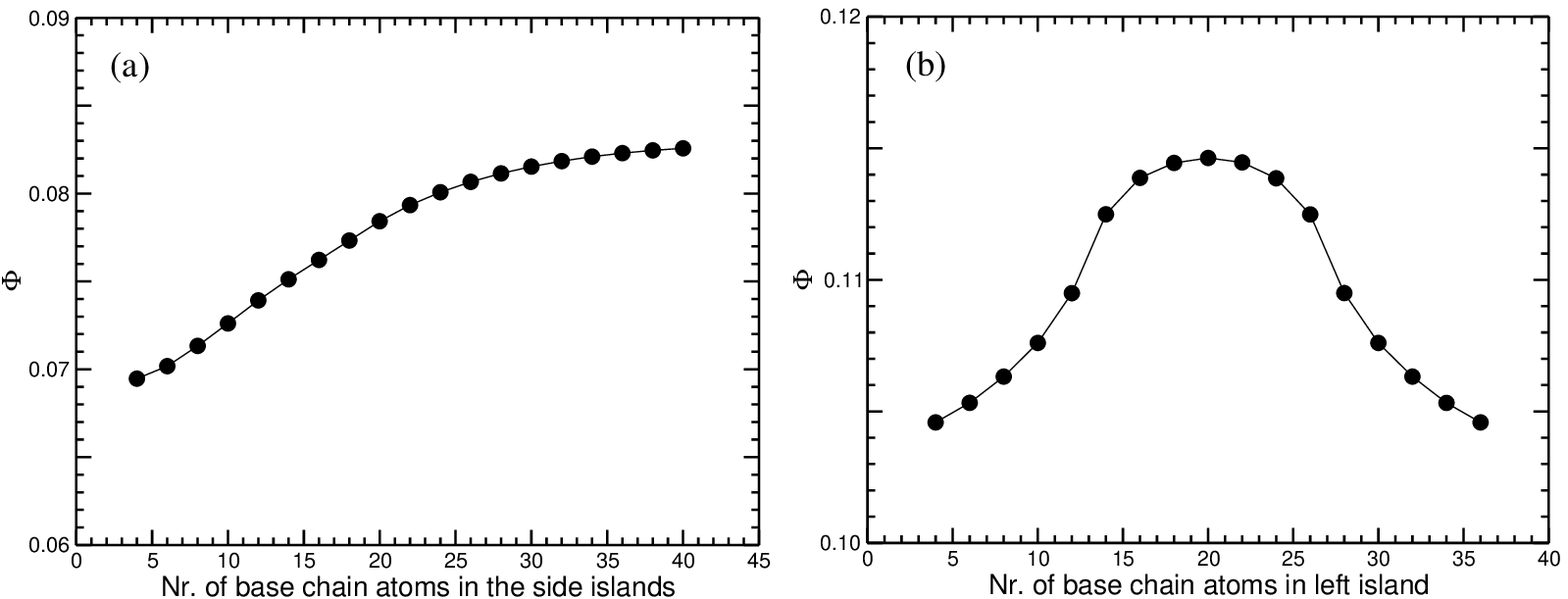}
\caption{\label{phiasym} \small Dependence of the wetting parameter
of the central island on the size of the two side islands for two
different configurations. In both cases all the islands are 3~ML
high, are separated by a distance $r$ = 5 and the central one has 20
atoms in the base chain. In (a) both side islands have equal volumes
and the misfit amounts to 7.0~\%. In (b), the sum of the volumes of
left and right islands is kept constant and equal to the doubled
volume of the central island; the misfit is 8.0~\%.}
\end{figure}

The size distribution of the neighbouring islands is expected to be
an important factor affecting the wetting of the island.
Figure~\ref{phiasym} shows the dependence of the wetting parameter
of the central island on the size of the side islands for two
different configurations. In the first [Figure~\ref{phiasym}(a)],
the two side islands have identical sizes. Increasing their volume
leads to an increase of the elastic fields around them and to a
further reduction of the bonding between the edge atoms of the
central island and the wetting layer. Figure~\ref{phiasym}(b) shows
the behaviour of the wetting parameter $\Phi $ of the central island
as a function of the number of atoms in the base chain of the left
island, whereby the sum of the total number of atoms of left and
right islands has been kept constant at precisely the doubled number
of the central island. The facet angles of all three islands are
always 60$^\circ$. Thus the first (and, by symmetry, also the last)
point give the maximum asymmetry in the size distribution of the
array, the left (right) island containing 9 atoms and the right
(left) island 105. All three islands are 3~ML thick. The point at
the maximal wetting corresponds to the monodisperse distribution:
i.e., when the three islands have one and the same volume of 57
atoms. This means that in the case of perfect self-assembly of the
array, the wetting parameter and therefore the tendency to
clustering display a maximum value.

\begin{figure}[htb]
\center
\includegraphics*[width=8.0cm]{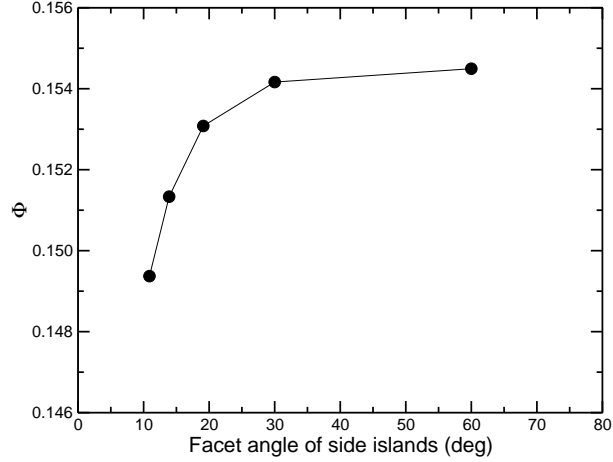}
\caption{\label{phivsalpha} \small Dependence of the wetting
parameter of the central island on the shape of the neighbouring
islands, as given by the facet angle of their side walls. The
central island has side walls of 60$^\circ$. All islands are 3~ML
high, have 20 atoms in their base chains and are separated by a
distance $r$~=~5. The misfit amounts to 8.0~\%.}
\end{figure}

As a further parameter affecting the wetting of an island, the
influence of the shape of the side islands, i.e. their facet angles,
on the wetting parameter of the central island is demonstrated in
Fig.~\ref{phivsalpha}. The central island is compact, having facet
angles of 60$^\circ$. The effect is greatest when the side islands
have the steepest walls. The same result (not shown) is obtained for
different shapes of the central island. The explanation follows the
same line as the one given above. The islands with larger-angle side
walls exert a greater elastic effect on the substrate and in turn on
the displacements and the bonding of the edge atoms of the central
island.

\begin{figure}[htb]
\center
\includegraphics*[width=8.0cm]{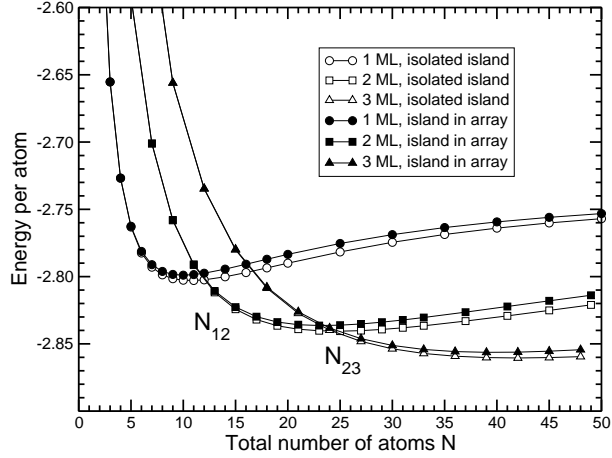}
\caption{\label{n123arr} \small Dependence of the energy per atom,
in units of V$_0$, on the total number of atoms in compressed
coherently strained islands with different thicknesses for a misfit
of 7.0~\%. The considered island has two identical neighbours at a
distance $n = 5$. The analogous curves for single isolated islands
(empty symbols) are also given for comparison. The wetting layer
consists in all cases of 3~ML which are allowed to relax.}
\end{figure}

The effect of the neighbours on the stability of islands with a
thickness increasing by one monolayer is illustrated in
Fig.~\ref{n123arr}. The total energy for islands inside an array
shows the same behaviour as for isolated islands. This means that
also in the case of a deformable substrate, the overall
transformation from the precursor 2D islands to the 3D islands takes
place in consecutive stages in each of which the islands thicken by
one monolayer. The energies computed in the case of the reference
single islands always lie below the curves of the islands in an
array. The difference obviously gives the energy of repulsion
between the neighbouring islands. It follows that the presence of
neighbouring islands leads to a slight decrease of the critical
misfit and in turn to a greater tendency to clustering.

\section{Discussion}

For the discussion of the above results we have to bear in mind that
a positive value of the wetting parameter implies in fact a tendency
of the deposit to form 3D clusters instead of a planar film.

We conclude that in the coherent SK growth the incomplete wetting is
due to the reduced adhesion of the edge atoms as a result of the
lattice misfit. This effect is the TDF for the coherent 3D
islanding. The interfacial energy of the boundary between the
wetting layer and the 3D crystallites is not zero and cannot be
neglected. This is confirmed by the effect of the anharmonicity of
the atomic interactions. The stronger repulsive forces between the
atoms in the compressed overlayers lead to larger lateral and
vertical displacements, and in turn, to weaker wetting in comparison
with films that are under tensile stress. Expanded islands adhere
stronger to the wetting layer and the TDF is very small to produce
3D islands. As a result, coherent SK growth in expanded films could
be expected at very (unrealistically) large absolute values of the
negative misfit. The latter, however, depends on the materials
parameters (degree of anharmonicity, strength of the chemical bonds,
etc.) of the particular system and cannot be completely ruled out.
In agreement with the predictions of our model, Xie {\it et al}.
found that Si$_{0.5}$Ge$_{0.5}$ 3D islands are formed on
Si$_{x}$Ge$_{1-x}$ relaxed buffer layers only under a compressive
misfit larger than 1.4~\%. Films under tensile stress were always
stable against 3D islanding.\cite{Xie}

A high TDF for coherent 3D islanding requires a large absolute value
of the misfit. The sharp increase of the TDF with the misfit in
compressed overlayers (Fig.~\ref{phivsf}) leads to the appearance
of a critical misfit beyond which 3D islanding is possible. The
existence of a critical misfit clearly shows that the origin of the
3D islanding in the coherent SK growth is the incomplete wetting,
which in turn is due to the atomic displacements near the islands
edges. Leonard {\em et al}.\cite{Leo} have successfully grown
quantum dots of In$_{x}$Ga$_{1-x}$As on GaAs(001) with $x = 0.5$
($\varepsilon _{0} \approx 3.6~\%$) but 60~\AA{} thick 2D quantum
wells at $x = 0.17$ ($\varepsilon _{0} \approx 1.2~\%$). Walther
{\it et al.}\cite{Walther} found that the critical In composition is
approximately $x \cong 0.25$, or $\varepsilon _{0} \cong 1.8~\%$. As
mentioned above, a critical misfit of 1.4~\% has been found by Xie
{\it et al.} upon deposition of Si$_{0.5}$Ge$_{0.5}$ films on
relaxed buffer layers of Si$_{x}$Ge$_{1-x}$ with varying
composition.\cite{Xie}

The increase of $N_{12}$ with decreasing absolute value of the
misfit (Fig.~\ref{n12vsf}) in overlayers under tensile stress is not
as sharp as in compressed ones and the critical behaviour is not
clearly pronounced in this particular case. Comparing
Figs.~\ref{phivsf} and \ref{n12vsf} clearly shows the reason for
this behaviour. Nevertheless, we could expect a critical-like
behaviour in experiments. Pinczolits {\it et al}.\cite{Pinc} have
found that deposition of PbSe$_{1-x}$Te$_{x}$ on PbTe(111) remains
purely two dimensional when the misfit is less than 1.6~\% in
absolute value (Se content $< 30 \%$).

In addition to the need of a sufficiently weak adhesion, the
existence of a critical misfit appears as a result of the
consideration of the island height as a discrete rather than a
continuous variable. The latter is also the reason for the complete
2D-3D transformation to pass through a series of intermediate states
with discretely increasing thickness. These states are
thermodynamically stable in consecutive intervals of the volume. The
first step of these transformations is the rearrangement of
monolayer-high islands into bilayer islands. Thus the 2D islands
appear as natural precursors for the formation of 3D islands beyond
some critical size. Moison {\it et al}. reported a sudden decrease
of the surface coverage from 1.75 to 1.2 ML at the moment of
formation of InAs 3D islands on GaAs.\cite{Moison} The same
phenomenon has been noticed by Shklyaev {\it et al}. in the case of
Ge/Si(111).\cite{Ichi} Voigtl\"ander and Zinner noted that Ge 3D
islands in Ge/Si(111) epitaxy have been observed at the same
locations where 2D islands locally exceeded the critical wetting
layer thickness of 2 bilayers\cite{Voigt1}. Ebiko {\it et al}. found
that the volume distribution of InAs/GaAs self-assembled quantum
dots agrees well with the scaling function characteristic of
monolayer-high islands in submonolayer epitaxy.\cite{Ebiko}

The question of the existence and particularly the stability of the
intermediate states is more difficult to answer. Rudra {\it et al.}
measured photoluminescence (PL) spectra of InAs layers deposited on
InP(001) at two different temperatures (490 and 525$^\circ$C) and
buried in the same material.\cite{Rudra} When the layers were grown
at 490$^\circ$C and the capping layer was deposited immediately
after the deposition of the InAs, the spectrum consisted of a single
line. If the InAs layer was annealed for 10~s before capping with
InP, the spectrum consisted of 8 lines. At 525$^\circ$C, 3 lines were
observed already in absence of annealing. The above observations
could be explained by the formation and coexistence of islands with
different thickness varying by one monolayer. Colocci {\it et
al.}\cite{Col} performed PL studies of InAs deposits on GaAs(001)
with thicknesses slightly varying around the critical thickness of
1.6~ML for the onset of the 3D islanding. They observed an
increasing number of luminescence lines with increasing film
thickness. These lines were attributed to families of 3D islands
with similar shape but with heights differing by one monolayer.

Although the above results seem to be in an excellent qualitative
agreement with the theoretical predictions of the model, the
thermodynamic stability of islands with quantized height of one
monolayer, and the existence of a critical misfit is still
debated.\cite{Politi,Duport} The reason of the discrepancy of our
results with those of Duport {\it et al.}\cite{Duport} most probably
stems from the implicit assumption, made by the above authors, that
the lengths of the lower ($R$) and the upper ($R^{\prime}$) bases,
and particularly the height ($h$) of the crystal they consider, which
has the shape of a frustum of a pyramid, are continuous variables. This
is correct if the crystals are sufficiently large. However, the
continuum approximation is not acceptable in the beginning of the
2D-3D transformation when the islands are still very small (and
thin). It is also not applicable in the limit $h \ll R$ for the same
reason. The problem is in fact analogous to the applicability of the
classical nucleation theory to small clusters. A cluster of a given
finite size is stable in a finite interval rather than for a fixed
value of the supersaturation. These intervals decrease with
increasing cluster size and become negligibly small when the
clusters are so large that the classical theory of nucleation
becomes valid.\cite{Mar2,Sasho} In the same way the continuum
approximation with $h \ll R$ should be applicable to much larger
islands consisting of several million of atoms.

Considering the self-assembly, the presence of neighbouring islands,
particularly of those highly compact (with largest angle facets), in
the near vicinity of a certain island decreases the adhesion of the
latter. The transformation of 2D, monolayer-high islands into
bilayer islands takes place by detachment of atoms from the edges
and their subsequent jumping and nucleating on the top island's
surface.\cite{StMar} This edge effect shows the influence of the
lattice misfit on the rate of second layer nucleation and in turn on
the kinetics of the 2D-3D transformation.\cite{Fil,Lin} The presence
of neighbouring islands favours the formation of 3D clusters and
their further growth. For the self-assembled monodisperse
population, the highest tendency to clustering is found.

We can regard the flatter islands in our model (11$^\circ$ facet
angle) as the ``hut" clusters discovered by Mo {\it et
al.},\cite{Mo} and the clusters with 60$^\circ$ facet angle as the
``dome" clusters. It is well known that clusters with steeper side
walls relieve the strain more efficiently than flatter
clusters~\cite{Mich} (a planar film, the limiting case of the flat
islands, does not relieve strain at all). We see that large-angle
facet islands affect more strongly the growth of the neighbouring
islands, leading to a more narrow size distribution.

Our results lead us to expect a self-assembled population of quantum
dots with highest density at low tem\-pe\-ra\-tures such that the
critical wetting layer thickness for 3D islanding approaches an
integer number of MLs. In InAs/GaAs quantum dots the reported values
of the critical thickness vary from 1.2 to 2~ML.\cite{Polimeni} The
critical wetting layer thickness is given by an integer number $n$
of MLs plus the product of the 2D island density and the critical
volume $N_{12}$ in the ($n$+1)-th~ML. The 2D island density
increases steeply with decreasing temperature.\cite{Joyce} In such a
case a dense population of 2D islands will overcome simultaneously
the critical size $N_{12}$ to produce bilayer islands. The value of
$N_{12}$ will be slightly reduced when neighbouring islands are
present due to the increase of $ \Phi $. Regions of high adatom
concentrations will favour the highest degree of self-assembly and,
due to the larger elastic forces present, are also likely to promote
the spatial ordering of the islands, possibly extending to less
dense regions and leading to self-organized arrays. Islands will
thus interact with each other from the very beginning of the 2D-3D
transformation and will give rise to the maximum possible wetting
parameter and, in turn, to islands with large-angle facets and a
narrow size distribution. This is in agreement with the observations
of self-assembled Ge quantum dots on Si(001).\cite{Letanh} At
700$^\circ$C a population of islands with a concentration ranging
from $10^7$ to $10^8$ cm$^{-2}$ is obtained. The islands have the
shape of truncated square pyramids with their side wall facets
formed by (105) planes (inclination angle of about 11$^\circ$). The
size distribution of the islands is quite broad. At 550$^\circ$C a
population of islands with an areal density of the order of $10^9$
to $10^{10}$ cm$^{-2}$ is observed. The islands have larger-angle
(113) facets and their size distribution is much more narrow.

Summarizing, the wetting layer and the 3D islands represent
different phases which cannot be in equilibrium with each other and
the SK morphology is a result of the replacement of one first order
phase transition (vapour -- wetting layer) by another first order
transition (vapour -- 3D islands). The transfer of mat\-ter from the
stable wetting layer to the 3D islands is thermodynamically
unfavoured. The experimental observations can be explained on the
base of two as\-sump\-tions: the thermodynamic driving force for the
coherent 3D islanding is the incomplete wetting and the height of
the 3D islands is a discrete variable varying by one monolayer. This
leads to the results that (i) monolayer-high islands with a critical
size appear as necessary precursors for 3D islands, (ii) the 2D-3D
transition takes place through a series of intermediate states with
discretely increasing thickness that are stable in separate
intervals of volume (iii) there exists a critical misfit below which
coherent 3D islands are thermodynamically unfavoured and the misfit
is accommodated by misfit dislocations at a later stage of the
growth. Compressed overlayers show a greater tendency to 3D
clustering than expanded ones, in agreement with experimental
results.

An open question is the mechanism by which the 3D nucleation takes
place on top of the wetting layer. It is well known that a
one-dimensional model is inadequate for the treatment of this case,
since it will always give a critical nucleus size of one
atom.\cite{Mar2} We are confident that in the future, a
generalization of our energy minimization model to 2+1 dimensions
will allow us to study the 3D nucleation problem in the coherent SK
growth.

Finally, the presence of neigh\-bouring is\-lands de\-creases the
wetting of the substrate (in this case the wetting layer) by the 3D
islands. The wetting decreases with increasing array density and
facet angle of the neighbouring islands. The wetting parameter
displays a maximum (implying a minimal wetting) when the array shows
a monodisperse size distribution. We expect an optimum
self-assembled islanding at low temperatures such that the 2D-3D
transformation takes place at the highest possible island density.

{\bf Acknowledgments}

J. E. P. gratefully acknowledges financial support from 
a ``Ram\'on y Cajal" contract of the Spanish MCyT.
E. K. is financially supported by the Spanish MCyT
Grant BFM 2001-291-C02-01 and Plan de Promoci\'on de la
Investigaci\'on UNED'02.

\end{document}